\documentclass[10pt,journal,doublecolumn,twoside]{IEEEtran}
\usepackage{}
\IEEEoverridecommandlockouts

\usepackage{subfigure}
\usepackage{colortbl}
\usepackage{bm}

\usepackage{graphicx}  
\usepackage{url}       
\usepackage{rotating}
\usepackage{amsmath}   
\usepackage{cite}

\usepackage{amsfonts,amssymb}

\usepackage{stfloats}

\usepackage{cases}
\usepackage{algorithm}
\usepackage{multirow}
\usepackage{algorithmic}

\newcommand{\myvec}[1]%
{\stackrel{\raisebox{-2pt}[0pt][0pt]
{\small$\rightharpoonup$}}{#1}}

\newcommand{\ls}[1]
    {\dimen0=\fontdimen6\the\font
     \lineskip=#1\dimen0
     \advance\lineskip.5\fontdimen5\the\font
     \advance\lineskip-\dimen0
     \lineskiplimit=.9\lineskip
     \baselineskip=\lineskip
     \advance\baselineskip\dimen0
     \normallineskip\lineskip
     \normallineskiplimit\lineskiplimit
     \normalbaselineskip\baselineskip
     \ignorespaces
    }

\usepackage{etoolbox}
\makeatletter
\patchcmd{\@makecaption}
  {\scshape}
  {}
  {}
  {}
\makeatother

\usepackage[absolute,showboxes]{textpos}
\TPMargin{3pt}

\begin{document}
\newcommand{\copyrightstatement}{
    \begin{textblock}{0.84}(0.08,0.95) 
         \noindent
         \footnotesize
         \copyright 2019 IEEE. Personal use of this material is permitted. Permission from IEEE must be obtained for all other uses, in any current or future media, including reprinting/republishing this material for advertising or promotional purposes, creating new collective works, for resale or redistribution to servers or lists, or reuse of any copyrighted component of this work in other works. DOI: 10.1109/GLOBECOM38437.2019.9013754.
    \end{textblock}
}
\copyrightstatement
\setlength{\columnsep}{0.24in}

\title{Mode Hopping With OAM-Based Index Modulation}

\author{Liping Liang$^{\dagger}$, Wenchi Cheng$^{\dagger}$, Wei Zhang$^{\ddagger}$, and Hailin Zhang$^{\dagger}$

\IEEEauthorblockA{$^{\dagger}$State Key Laboratory of Integrated
Services Networks, Xidian University, Xi'an, China\\
$^{\ddagger}$ University of New South Wales, Sydney, Australia\\
E-mail: \{\emph{lpliang@stu.xidian.edu.cn}, \emph{wccheng@xidian.edu.cn}, \emph{wzhang@ee.unsw.edu.au}, \emph{hlzhang@xidian.edu.cn}\}}

\thanks{This work was supported in part by the National Natural Science Foundation of China (Nos. 61771368 and 61671347), Young Elite Scientists Sponsorship Program By CAST (2016QNRC001), the 111 Project of China (B08038), Doctoral Student's Short Term Study Abroad Scholarship Fund of Xidian University, and the Australian Research Council's Projects funding scheme under Projects (DP160104903 and LP160100672). }

}

\maketitle
\thispagestyle{empty}
\pagestyle{empty}

\begin{abstract}

Orbital angular momentum (OAM) based mode hopping (MH) scheme is expected to be a potential anti-jamming technology in radio vortex wireless communications. However, it only uses one OAM-mode for hopping, thus resulting in low spectrum efficiency (SE). Index modulation offers a trade-off balance between the SE and performance reliability. In this paper, we propose an MH with OAM-based index modulation scheme, where several OAM-modes are activated for hopping, to achieve high SE at a given bit error rate in radio vortex wireless communications. Based on the proposed scheme, we derive the upper bound and lower bound of achievable SEs. Furthermore, in order to take advantage of index information, we derive the optimal hopped OAM-modes to achieve the maximum SE. Numerical results show that our proposed MH with index modulation scheme can achieve high SE while satisfying a certain reliability of radio vortex wireless communications.

\end{abstract}


\section{Introduction}

\IEEEPARstart{E}{merging} radio vortex wireless communication is expected to solve the spectrum shortage problem caused by the growing traffic data and multiple serves for the future wireless communications~\cite{OAM_mag,OAM_study}. Radio vortex wireless communications take advantage of orbital angular momentum (OAM), a category of angular momentum, for transmission. OAM waves generated by many methods, such as uniform circular array (UCA) and spiral phase plate (SPP), are overlapping but orthogonal with each other~\cite{2013_axis,2011_Eletromagnetic}. Hence, signals with OAM-modes can be transmitted without inter-mode interference. 

Recently, OAM applications have been extensively studied in radio vortex wireless communications, such as OAM-based multiple-input multiple-output (MIMO), OAM multiplexing jointly used with the traditional orthogonal frequency division multiplexing (OFDM) for high spectrum efficiency (SE), radar imaging, OAM waves converging, microwave sensing, and mode hopping (MH) for anti-jamming~\cite{OAM_MIMO,OAM_OFDM,OAM_imaging,len_2018,OAM_sensing,MH}. The OAM-based wireless channel model was built and the OAM signals were decomposed in sparse multipath environments containing a line-of-sight (LoS) path and several reflection paths~\cite{OAM_OFDM}.
In OAM-based MH scheme, only one OAM-mode is activated for hopping at each time-slot~\cite{MH}. Thus, if the number of OAM-modes is relatively large, most OAM-modes cannot work, thus resulting in the resource waste. Also, without using multiple OAM-modes for transmission, the achievable SE of proposed MH scheme is low. 

Index modulation, an emerging concept, is the extension of spatial modulation used in MIMO communications~\cite{SM,index}. Index modulation activates antennas of MIMO system or subcarriers of OFDM system, to convey signal information. The indices of activated antennas or subcarriers, which can be considered as an information source, convey the additional information, thus increasing the achievable SE of wireless communications. 
It was recently verified that OAM-based index modulation scheme without increasing the size of signal constellation can achieve better error performance than the conventional OAM multiplexing scheme without increasing the detection complexity~\cite{OAM_index}.
However, how to take advantage of the maximum index information in MH scheme to achieve high SE while satisfying a certain reliability still is an open challenge.

In this paper, we propose an MH with OAM-based index modulation scheme, where some OAM-modes are selected to hop for anti-jamming, to achieve high SE at a given error performance in radio vortex wireless communications. The main idea is to randomly activate some OAM-modes for each hop to increase the SE and OAM-modes utilizing efficiency. 
Based on the proposed scheme, the upper and lower bound of SEs for each hop are derived, respectively. The bounds of SEs are obtained for a given number of hopped OAM-modes. Then, we analyze the relationship between signal-to-noise ratio (SNR) and the number of hopped OAM-modes and derive the optimal solution of hopped OAM-modes at low channel SNR region. Numerical results show that
our proposed MH with index modulation scheme outperforms the conventional OAM multiplexing scheme at low SNR region.

\begin{figure*}
\centering
  \includegraphics[width=1.02\textwidth]{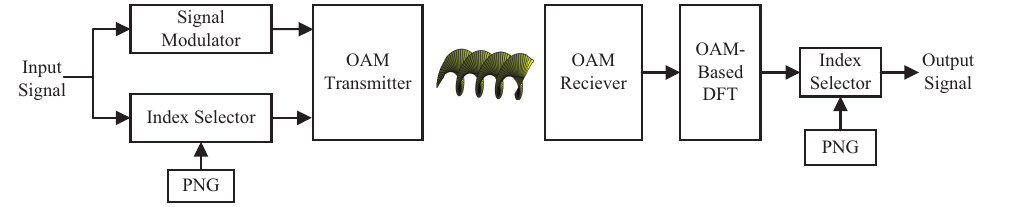}\\
  \caption{The system modulation of MH with OAM-based index modulation in radio vortex wireless communications.}\label{fig:sys}
\end{figure*}

The remainder of this paper is organized as follows. Section~\ref{sec:sys} gives the MH with OAM-based index modulation system model. Section~\ref{sec:index} presents the OAM-based index modulation scheme, derives the upper lower bounds of the achievable SE, and calculates how many hopped OAM-modes should be selected to achieve higher SE. Section~\ref{sec:performance} evaluates the MH with OAM-based index modulation scheme and compares it with the conventional OAM multiplexing scheme. The paper concludes with Section~\ref{sec:conc}.

\section{System Model}\label{sec:sys}
In this section, we build the system model of MH with OAM-based index modulation as shown in Fig.~\ref{fig:sys}. In this paper, we use the UCA as the OAM-transmitter and OAM-receiver to generate and receive multiple OAM-modes, respectively. There are $N_{t}$ and $N_{r}$ element-arrays, which are equidistantly distributed around the perimeter of the circle, for the transmit and receive UCAs, respectively. The $N_{t}$ element-arrays are fed with the same input signals, but with a successive delay from element to element such that after a full turn the phase has been incremented by an integer multiple $l$ of $2\pi$, where $l$ is the order of OAM-modes and satisfies $|l|\leq N_{t}/2$~\cite{2011_Eletromagnetic}.

As shown in Fig,~\ref{fig:sys}, the transmit data is split into index selector and signal modulator. The index selector is controlled by pseudo-noise generator (PNG). With index selector, some OAM-modes are activated for hopping and the other OAM-modes are inactivated. The activated OAM-modes are selected by the index selector, which selects $I$ OAM-modes out of $N_{t}$ OAM-modes for signal transmission. It means that there are $I$ OAM-modes hopping simultaneously for each time-slot. Thus, there are $K={N_{t}\choose I}$ combinations for OAM signal transmission. We assume that the set of selected OAM-modes for transmission is $L$. Thus, the corresponding $k$-th ($1 \leq k \leq K$) combination is denoted by $L_{k}$. The input signal is conveyed by the activated multiple OAM-modes with signal transmission and the indices of OAM-modes with the combination of OAM-modes. The inactivated OAM-modes don't work.

\begin{figure}
\centering
  \includegraphics[width=0.33\textwidth]{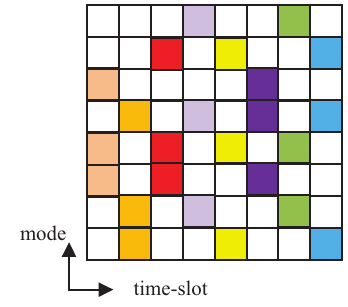}\\
  \caption{MH pattern with OAM-based index modulation.}\label{fig:MH}
\end{figure}

At the receiver, the OAM signals can be decomposed by the discrete Fourier transform (DFT) algorithm. Note that PNGs at the transmitter and receiver are same. In order to filter the interference OAM signals, the decomposed signals are selected by the index selector. We assume that the signal can be transmitted by one hop. Also, for utilizing OAM multiplexing, the signals with different hopped OAM-modes are different. An example of MH pattern with OAM-based index-modulation scheme is presented in Fig.~\ref{fig:MH}, where the specified color identifies each hop, $N_{t}=8$, and $I=3$.

With the proposed index modulation, wireless communication can not only achieve high SE, but also satisfy the reliability. In the following, we mainly focus on analyzing the proposed MH with OAM-based index modulation for high achievable SE in radio vortex wireless communications.

\section{OAM-Based Index Modulation}~\label{sec:index}
In this section, we propose the MH with OAM-based index modulation scheme in radio vortex wireless communications. Then, we derive the closed-form expressions of lower bound and upper bound of achievable SEs, respectively. We also analyze how many OAM-modes should be selected to obtain higher SE.

\subsection{Index Modulation}

In the subsection, we consider the $k$-th combination of $I$. We assume that the set of selected OAM-modes for transmission is $L_{k}=\{l_{1,k},\cdots, l_{i,k}, \cdots, l_{I,k}\}$, where $l_{i,k} (1 \leq i \leq I)$ is the $i$-th activated OAM-mode of the all $I$ OAM-modes. Also, we assume that the set of $L$ is arranged in the order of OAM-modes, that is, $l_{1,k}< \cdots < l_{i,k}< \cdots < l_{I,k}$. For example, if the hopped OAM-mode $l_{1,k}=2$, it means OAM-mode $l=2$ is activated.

In this paper, we only consider a user for transmission without other interference users. We assume that the vector of transmit modulated signal is $\bm{S}_{k}=[s_{1,k},\cdots, s_{i,k}, \cdots, s_{I,k}]^{T}$ for given set $L_{k}$, where $s_{i,k}$ is the $i$-th transmit signal for the $l_{i,k}$ OAM-modes and $[\cdot]^{T}$ is the transpose operation. The inactivated OAM-modes transmit zero signals with no power. Thus, the vector of transmit signals can be expressed as $\bm{s}_{k}=[0, \bm{S}_{k}]^{T}$, which contains $(N_{t}-I)$ zeros and $I$ transmit signals. The positions of zero terms is determined by the selected activated OAM-modes.

Therefore, the transmit OAM signal, denoted by $s_{n,i,k}$, for the $n$-th $(0 \leq n \leq N_{t}-1)$ element-array corresponding to the $i$-th hopped OAM can be expressed as follows:
\begin{equation}
    s_{n,i,k}=s_{i,k} e^{j\frac{2\pi n}{N_{t}}l_{i,k}}.
\end{equation}
The transmit signal, denoted by $x_{n,k}$, for the $n$-th transmit element-array can be derived as follows:
\begin{equation}
    x_{n,k}=\sum_{i=1}^{I} s_{n,i,k}=\sum_{i=1}^{I} s_{i,k} e^{j\frac{2\pi n}{N_{t}} l_{i,k}}.
    \label{ea:x_n}
\end{equation}

At the receiver, the received signal, denoted by $\bm{r}_{k}$, for the $N_{r}$ element-arrays can be obtained as follows:
\begin{equation}
    \bm{r}_{k}=\bm{H}\bm{x}_{k}+\bm{n}_{k},
    \label{eq:rec_y}
\end{equation}
where $\bm{H}$ is the $N_{r}\times N_{t}$ channel gain matrix, $\bm{x}_{k}$ is the vector of transmit signal with respect to $x_{n,k}$, and $\bm{n}_{k}$ is the Gaussian noise received at the receiver.

According to Eq.~\eqref{ea:x_n}, the expression of Eq.~\eqref{eq:rec_y} can be re-expressed as follows:
\begin{equation}
    \bm{r}_{k}= \bm{HF}\bm{s}_{k}+\bm{n}_{k},
\end{equation}
where $\bm{F}$ is a $N_{t}\times N_{t}$ inverse discrete Fourier transform (IDFT) matrix.

To decompose the OAM signal, the DFT algorithm is used at the receiver. Then, we have
\begin{eqnarray}
   \tilde{\bm{r}}_{k}&=& \bm{F}^{H}\bm{HF}\bm{s}_{k}+\tilde{\bm{n}}_{k}
    \nonumber\\
    &=& \widetilde{\bm{H}}\bm{s}_{k}+\tilde{\bm{n}}_{k},
\end{eqnarray}
where $\tilde{\bm{r}}_{k}$ is denoted by the vector of decomposed OAM signals, $(\cdot)^{H}$ represents the conjugate transpose operation and $\tilde{\bm{n}}_{k}$ is the noise after DFT algorithm. $\widetilde{\bm{H}}=\text{diag}\{h_{-N_{t}/2+1},\cdots, h_{l}, \cdots, h_{N_{t}/2}\}$ is the channel gain diagonal matrix with respect to each OAM-mode, where $h_{l}$ is denoted by the channel gain for the $l$-th OAM-mode from the OAM-transmitter to OAM-receiver.

For an UCA-based transceiver in LoS transmission, the channel gain $h_{l}$ can be derived as~\cite{MH}
\begin{equation}
    h_{l}\!=\!\frac{\beta \lambda N_{t} j^{-l}e^{-j\frac{2\pi}{\lambda} \sqrt{D^{2}+R_{1}^{2}+R_{2}^{2}}}}{4\pi \sqrt{D^{2}+R_{1}^{2}+R_{2}^{2}}}\!  e^{j\varphi l}\! J_{l}\!\!\left(\!\frac{2\pi R_{1} R_{2}}{\lambda \sqrt{D^{2}\!+\!R_{1}^{2}+R_{2}^{2}}}\! \right),
\end{equation}
where $\beta$ contains all relevant constants such as attenuation and phase rotation caused by antennas and their patterns on both sides, $\lambda$ represents the carrier wavelength, $D$ is denoted by the distance from the OAM-transmitter center to the OAM-receiver center, $R_{1}$ is the radius of OAM-transmitter, $R_{2}$ is the radius of OAM-receiver, and $J_{l}(\cdot)$ is the $l$-th order of Bessel function.
Hence, we can obtain the corresponding $h_{i,k}$ for the $i$-th hopped OAM-mode.

However, using DFT algorithm can decompose all OAM signals for activated and inactivated OAM-modes, thus interfering the useful signals. Take example of $L_{k}=\{-1, 0, 2\}$, where $I$ is set as $3$. The indices of decomposed signals are $\{-1,0,1,2\}$, which implies 1 OAM signal is from interference. Thus, the index selector is required to filter interference which carries different OAM-modes from the hopping modes and select the wanted signals. Therefore, the received signal, denoted by ${\bm{y}}_{k}$, can be expressed as follows:
\begin{equation}
    {\bm{y}}_{k}= \Lambda \bm{S}_{k}+ \bm{w}_{k},
\end{equation}
where $\bm{w}_{k}$ is the vector of received noise with respect to the activated OAM-modes and $\Lambda=\text{diag}\{h_{0,,k}, \cdots,h_{i,k},\cdots,h_{I,k}\}$.

\subsection{Spectrum Efficiency}

The transmit power of signal is denoted by $P_{i,k}$ for the $i$-th hopped OAM-mode. Thus, the variance, denoted by $\bm{\Sigma}_{s,k}$, of $\bm{S}_{k}$ can be expressed as follows:
\begin{eqnarray}
    \bm{\Sigma}_{s,k}=\text{diag}\{ P_{1,k},\cdots, P_{i,k},  \cdots, P_{I,k}  \}.
\end{eqnarray}
We also have the noise variance, denoted by $\bm{\Sigma}_{w,k}$, as follows:
\begin{eqnarray}
    \bm{\Sigma}_{w,k}=\text{diag}\{\sigma_{1,k}^{2}, \cdots, \sigma_{i,k}^{2}, \cdots, \sigma_{I,k}^{2}\},
\end{eqnarray}
where $\sigma_{i,k}^{2}$ is denoted by the noise variance for the $i$-th hopped OAM-mode.

Since the transmit signal and noise follow multivariate normal distribution, the received signal $\bm{{y}}_{k}$ also follows and the corresponding variance, denoted by $\bm{\Sigma}_{y,k}$, for the received signal can be expressed as follows:
\begin{eqnarray}
    \bm{\Sigma}_{y,k}=\Lambda \bm{\Sigma}_{s,k} \Lambda^{H} + \bm{\Sigma}_{w,k}.
\end{eqnarray}
Thus, the conditional probability density function (PDF), denoted by $f(\bm{y}_{k})$, of $\bm{{y}}_{k}$ can be given as follows:
\begin{eqnarray}
    f(\bm{y}_{k})&=&\frac{1}{\pi^{N_{t}}|\bm{\Sigma}_{y,k}|} \exp\left(-\bm{y}^{H}\bm{\Sigma}_{y,k}^{-1}\bm{y}\right).
    \label{eq:f_y}
\end{eqnarray}

To obtain the achievable SE of our proposed MH with index modulation scheme in radio vortex wireless communications, mutual information is used. Based on the analysis mentioned above, the mutual information, denoted by $I(\bm{s}, \bm{y})$, between the transmit signal vector $\bm{S}_{k}$ and received signal vector ${\bm{{y}}_{k}}$ is given as follows~\cite{tse2005fundamentals}:
\begin{eqnarray}
    I(\bm{s}, \bm{y})= I(\bm{s}_{k},\bm{y}| X_{k}^{I})+I( X_{k}^{I},\bm{y}),
    \label{eq:eq:I_sy}
\end{eqnarray}
where $I(\bm{s}_{k},\bm{y}| X_{k}^{I})$ is the signal information, $I( X_{k}^{I},\bm{y})$ is the index information, $X_{k}^{I}$ represents the index information, and $\bm{y}$ is the decomposed OAM signal vector for all possible $k$. The first term on the right-hand of Eq.~\eqref{eq:eq:I_sy} can be derived as follows:
\begin{eqnarray}
    I(\bm{s}_{k},\bm{y}| X_{k}^{I})&=& \frac{1}{K}\sum_{k=1}^{K}\log_{2}\frac{|\bm{\Sigma}_{y,k}|}{|\bm{\Sigma}_{w,k}|}\nonumber\\
    &=&\frac{1}{K}\sum_{k=1}^{K}\sum_{i=1}^{I}\log_{2}(1+P_{i,k}\gamma_{i,k}),
\end{eqnarray}
where $|\cdot|$ represents the determinant of matrix and $\gamma_{i,k}=h_{i,k}^{2}/\sigma_{i,k}^{2}$.

Then, the index information $I( X_{k}^{I},\bm{y})$ can be derived as follows:
\begin{eqnarray}
    I( X_{k}^{I},\bm{y})= H(\bm{y})-H(\bm{y}|\bm{s}_{k}, X_{k}^{I}),
\end{eqnarray}
where $H(\bm{y})$ and $H(\bm{y}|\bm{s}_{k}, X_{k}^{I})$ represent the entropy and conditional entropy of the received signal, respectively.

The PDF, denoted by $f(\bm{y})$, for all possible $k$ can be derived as follows:
\begin{eqnarray}
    f(\bm{y})=\frac{1}{K}\sum_{k=1}^{K}f(\bm{y}_{k}).
\end{eqnarray}
Based on Jensen's inequality, entropy of received signal $H(\bm{y})$ can be derived as follows:
\begin{eqnarray}
   H(\bm{y})\hspace{-0.2cm} &=& \hspace{-0.2cm}-\int_{\bm{y}} f(\bm{y})\log_{2}f(\bm{y})\text{d}\bm{y}
   \nonumber\\
  \hspace{-0.2cm} &\geq &\hspace{-0.2cm}- \frac{1}{K}\sum_{k=1}^{K}\log_{2}\left[\frac{1}{K}\sum_{j=1}^{K} \int_{\bm{y}} f(\bm{y}_{j})f(\bm{y}_{k})\text{d}\bm{y}\right]
   \nonumber\\
   \hspace{-0.2cm}&=&\hspace{-0.2cm}- \frac{1}{K}\sum_{k=1}^{K}\log_{2}\left[\frac{1}{K}\sum_{j=1}^{K} \frac{\left|\bm{\Sigma}_{y,k}^{-1}+\bm{\Sigma}_{y,j}^{-1}\right|^{-1}}{\pi^{N_{t}} |\bm{\Sigma}_{y,k}||\bm{\Sigma}_{y,j}|}\right]
   \nonumber\\
   \hspace{-0.2cm} &=&\hspace{-0.3cm}- \frac{1}{K}\sum_{k=1}^{K}\log_{2}\left[\frac{1}{K}\sum_{j=1}^{K} \frac{1}{\pi^{N_{t}} |\bm{\Sigma}_{y,k}\!+\!\bm{\Sigma}_{y,j}|}\right].
\end{eqnarray}
For $H(\bm{y}|\bm{s}_{k}, X_{k}^{I})$, we have
\begin{eqnarray}
    H(\bm{y}|\bm{s}_{k}, X_{k}^{I})\hspace{-0.3cm}&=&\hspace{-0.3cm}
   \frac{1}{K}\!\!\sum_{k=1}^{K}\!\sum_{i=1}^{I}\!\log_{2}\left[\sigma_{i,k}^{2}(1\!+\!P_{i,k}\gamma_{i,k})\right] \!\!+\!\! I\log_{2}(\pi e).\nonumber\\
\end{eqnarray}
Therefore, the lower bound, denoted by $C_{low}$, of SE for our proposed MH with OAM-based index modulation scheme can be derived as follows:
\begin{eqnarray}
    C_{low}\hspace{-0.3cm}&=&\hspace{-0.3cm}\log_{2}K-\frac{1}{K}\sum_{k=1}^{K}\sum_{i=1}^{I}\log_{2}\sigma_{i,k}^{2} - I\log_{2} (\pi e)
    \nonumber\\
    &&\hspace{-0.3cm}-\frac{1}{K}\log_{2}\left[\prod_{k=1}^{K}\left(\sum_{j=1}^{K} \frac{1}{\pi^{I} |\bm{\Sigma}_{y,k}+\bm{\Sigma}_{y,j}|}\right)\right].
    \label{eq:C_LOW}
\end{eqnarray}
The expression of $I( X_{k}^{I},\bm{y})$ can be re-written as follows:
\begin{eqnarray}
    I( X_{k}^{I},\bm{y})=\frac{1}{K}\sum_{k=1}^{K} D\left[f(\bm{y}_{k})\parallel f(\bm{y})\right],
\end{eqnarray}
where $D\left[f(\bm{y}_{k})\parallel f(\bm{y})\right]$ is the Kullback-Leibler (KL) divergence between the Gaussian distribution with PDF $f(\bm{y}_{k})$ and Gaussian mixture with the PDF $f(\bm{y})$.

In the following, $(\cdot)^{-1}$ and $\text{Tr}(\cdot)$ represent the inverse operation and trace of the matrix. According to the approximative analysis~\cite{Approximating_KL}, the upper bound of KL divergence, denoted by $D_{up}\left[f(\bm{y}_{k})\parallel f(\bm{y})\right]$, can be derived as follows:
\begin{equation}
    D_{up}\left[f(\bm{y}_{k})\parallel f(\bm{y})\right]=-\log_{2}\left\{\sum\limits_{j=1}^{K} \frac{1}{K} e^{-D[f(\bm{y}_{k})\parallel f(\bm{y}_{j})]}\right\},
    \label{eq:dkl}
\end{equation}
where
    \begin{eqnarray}
     D\!\left[f(\bm{y}_{k})\parallel f(\bm{y}_{j})\right]\!&=&\!\log_{2} \frac{|\bm{\Sigma}_{y,j}|}{|\bm{\Sigma}_{y,k}|}-I\log_{2}e \nonumber\\
    &&\!\!\!+\log_{2}e{\bf{\bm{Tr}}}\left(\bm{\Sigma}_{y,j}^{-1}\bm{\Sigma}_{y,k}\right).
\end{eqnarray}
Therefore, the upper bound of achievable SE, denoted by $C_{up}$, with our proposed OAM-based index modulation in radio vortex wireless communications can be derived as follows:
\begin{eqnarray}
    C_{up}\hspace{-0.3cm}&=&\hspace{-0.3cm}\frac{1}{K}\sum_{k=1}^{K}\sum_{i=1}^{I}\log_{2}(1+P_{i,k}\gamma_{i,k})\nonumber\\
    &&\hspace{-0.3cm}-\frac{1}{K}\sum_{k=1}^{K}\log_{2}\left(\sum\limits_{j=1}^{K} \frac{1}{K} e^{-D[f(\bm{y}_{k})\parallel f(\bm{y}_{j})]}\right).
   \label{eq:C_UP}
\end{eqnarray}

\subsection{How Many OAM-Modes are Selected for Higher SE?}\label{sec:How}

For a fixed $I$, the upper and lower bounds of achievable SE for our proposed MH with OAM-based index modulation scheme in radio vortex wireless communications can be derived as shown in Eqs.~\eqref{eq:C_UP} and \eqref{eq:C_LOW}. However, how many OAM-modes should be selected to achieve higher SE for the proposed scheme? In the following, we will answer this question.

The expression of Eq.~\eqref{eq:dkl} can be re-written as follows:
\begin{eqnarray}
  &&\hspace{-1.2cm}D_{up}\left[f(\bm{y}_{k})\parallel f(\bm{y})\right]\nonumber\\
  &&\hspace{-1.2cm}=\log_{2}K-\frac{1}{K}\sum_{k=1}^{K} \log_{2}\left\{1+\sum_{j=1 \atop j\neq k}^{K}e^{-D\left[f(\bm{y}_{k})\parallel f(\bm{y}_{j})\right]}\right\}.
  \label{eq:dkl_re}
\end{eqnarray}
Since the KL divergence is always non-negative, $\log_{2}\left\{1+\sum_{j=1 \atop j\neq k}^{K}e^{-D\left[f(\bm{y}_{k})\parallel f(\bm{y}_{j})\right]}\right\}$ is non-negative. Thus, neglecting the second term on the right-hand of Eq.~\eqref{eq:dkl_re} yields the desired upper bound. Thus, the upper bound of SE $C_{up}$ can be re-expressed as follows:
\begin{eqnarray}
  C_{up}\hspace{-0.3cm}&=&\hspace{-0.3cm}\frac{1}{K}\sum_{k=1}^{K}\sum_{i=1}^{I}\log_{2}(1+P_{i,k}\gamma_{i,k})+\log_{2}K.
  \label{eq:c_up_re}
\end{eqnarray}
Actually, the upper bound given in Eq.~\eqref{eq:c_up_re} is significantly close to the true entropy value~\cite{KL_2008}. We assume that the transmit power for each hopped OAM-mode is $P_{0}$. The variance of the $l$-th OAM-mode is denoted by $\sigma_{l}^{2}$. For all possible $k$, each OAM-mode is hopped for $\binom {N_{t}-1} {I-1}$ times.
Thus, the first term on the right-hand of Eq.~\eqref{eq:c_up_re} can be re-expressed as follows:
\begin{eqnarray}
    \frac{1}{K}\sum_{k=1}^{K}\sum_{i=1}^{I}\log_{2}(1+P_{i,k}\gamma_{i,k})=\frac{I}{N_{t}}\sum_{l=0}^{N_{t}-1}\log_{2}(1+P_{0}\gamma_{l}),
\end{eqnarray}
where $\gamma_{l}=h_{l}^{2}/\sigma_{l}^{2}$. Thus, the upper bound of SE can be re-expressed as follows:
\begin{equation}
 C_{up}(I)=\frac{I}{N_{t}}\sum_{l=0}^{N_{t}-1}\log_{2}(1+P_{0}\gamma_{l})+\log_{2}\binom {N_{t}} {I}.
\end{equation}
Then, we denote by $f(z)$ the continuous function of $z$, which is defined as
\begin{eqnarray}
    f(z)\hspace{-0.2cm}&=& \hspace{-0.2cm} \frac{z}{N_{t}}\sum_{l=0}^{N_{t}-1}\log_{2}(1+P_{0}\gamma_{l}) +  \log_{2}\binom {N_{t}} {z}\nonumber\\
  \hspace{-0.2cm}  &=& \hspace{-0.2cm}  \frac{z}{N_{t}}\sum_{l=0}^{N_{t}-1}\log_{2}(1+P_{0}\gamma_{l}) + \log_{2}\Gamma(N_{t}+1)\nonumber\\
  &&\hspace{-0.3cm}-\log_{2}\Gamma(z+1)-\log_{2}\Gamma(N_{t}-z+1),
\end{eqnarray}
where $\Gamma(\cdot)$ is the Gamma function. We denote by $\mathcal {Q}(z)$ the derivative of function $\ln \Gamma(z)$. Thus, we have
\begin{eqnarray}
    \mathcal{Q}(z+1) =  \frac{1}{z}+\mathcal{Q}(z), \label{eq:Q}
\end{eqnarray}
where we use the characteristic given by
\begin{eqnarray}
    \Gamma(z)=z \Gamma(z).
\end{eqnarray}
Then, the derivative, denoted by $\mathcal{P}(N_{t}-z+1)$, of function $\ln \Gamma(N_{t}-z+1)$ can be calculated as follows:
\begin{eqnarray}
    \mathcal{P}(N_{t}-z+1)= - \mathcal{Q}(N_{t}-z+1).
\end{eqnarray}
Thus, the derivative, denoted by $f^{\prime}(z)$, of $f(z)$ can be written as follows:
\begin{equation}
    f^{\prime}(z)=\frac{1}{N_{t}}\sum_{l=0}^{N_{t}-1}\log_{2}(1+P_{0}\gamma_{l}) -\frac{\mathcal{Q}(z\!+\!1)}{\ln 2} +\frac{\mathcal{Q}(N_{t}\!-\!z\!+\!1)}{\ln 2}.
\end{equation}
Let $z=I$. Based on Eq.~\eqref{eq:Q}, we have
\begin{eqnarray}
    \mathcal{Q}(I+1)=\frac{1}{I}+\mathcal{Q}(I)
    =\sum_{i=1}^{I} \frac{1}{i}-\zeta,
\end{eqnarray}
where $\zeta=\mathcal{Q}(1)$ is the Euler-Mascheroni constant. The corresponding derivative of $f(I)$ can be re-expressed as follows:
\begin{eqnarray}
    f^{\prime}(I)\hspace{-0.3cm}&=&\hspace{-0.3cm}\frac{1}{N_{t}}\!\sum_{l=0}^{N_{t}-1}\!\log_{2}(1+P_{0}\gamma_{l}) \! -\!\frac{\mathcal{Q}(I\!+1)}{\ln 2} \!+ \frac{\mathcal{Q}(N_{t}-I+1)}{\ln 2} \nonumber\\
   \hspace{-0.3cm}& =& \hspace{-0.3cm}\frac{1}{N_{t}}\sum_{l=0}^{N_{t}-1}\log_{2}(1+P_{0}\gamma_{l}) -\frac{1}{\ln 2}\left[\sum_{i=1}^{I}\frac{1}{i}-\sum_{u=1}^{N_{t}-I}\frac{1}{u}\right].\nonumber\\
\end{eqnarray}
Clearly, $f^{\prime}(I) >0 $, when $I \leq \lfloor N_{t}/2 \rfloor$, where $\lfloor\cdot\rfloor$ is the floor function.
Then, we can find a value $I_{0}$, which satisfies $f^{\prime}(I_{0})\geq 0$ and $f^{\prime}(I_{0}+1)< 0$ with method of an exhaustive search.

If the channel SNR is relatively larger, the signal information mainly impacts the upper bound of SE in MH scheme. Thus, the the SE always increases as the number of hopped OAM-modes. However, when channel SNR is smaller, the index modulation plays the important role.  Hence, we have the following Lemma 1.

\emph{Lemma 1}: At low SNR region, the maximum SE of our proposed scheme, denoted by $C^{*}$, is
\begin{eqnarray}
    C^{*}=\max\{C_{up}(I_{0}),C_{up}(I_{0}+1)\},
\end{eqnarray}
where $I_{0}$ satisfies
\begin{eqnarray}
    \begin{cases}
       \ln 2 \frac{1}{N_{t}}\sum\limits_{l=0}^{N_{t}-1}\log_{2}(1+P_{0}\gamma_{l}) \geq \sum\limits_{i=N_{t}-I_{0}+1}^{2I_{0}-N_{t}}\frac{1}{i};\\
       \ln 2 \frac{1}{N_{t}}\sum\limits_{l=0}^{N_{t}-1}\log_{2}(1+P_{0}\gamma_{l})< \sum\limits_{i=N_{t}-I_{0}}^{2I_{0}+2-N_{t}}\frac{1}{i}.
    \end{cases}
\end{eqnarray}

\section{Performance Evaluations}\label{sec:performance}

In this section, we evaluate the performance of our developed MH with OAM-based index modulation scheme in radio vortex wireless communications. In the evaluation, the lens are used to converge OAM beams~\cite{len_2018}. Throughout our evaluations, we set the carrier frequency as 60 GHz, transmit distance as 3 m, and transmit power of each OAM-mode as 1 W.

\begin{figure}
\centering
  \includegraphics[width=0.53\textwidth]{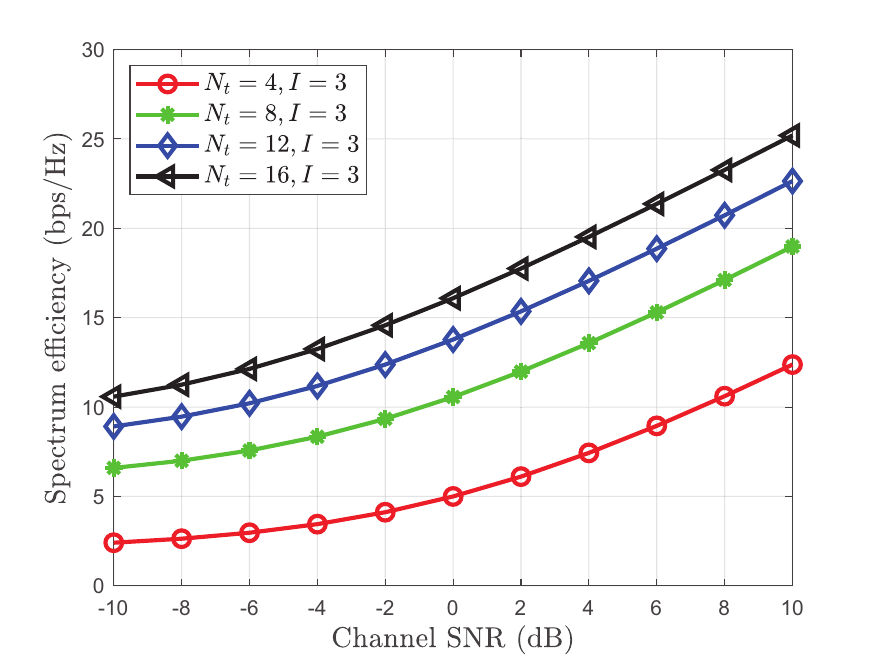}\\
  \caption{The spectrum efficiencies of different number of total OAM-modes for our proposed scheme.}
  \label{fig:SE_N}
\end{figure}

Figure~\ref{fig:SE_N} depicts the SEs of our proposed MH with OAM-based index modulation scheme versus the channel SNR in radio vortex wireless communications, where we set $I=3$, $N_{t}=4,8,12$ as well as 16, respectively. As shown in Fig.~\ref{fig:SE_N}, for a given number of hopped OAM-modes, the SEs of our proposed scheme increases as the number of total OAM-modes increases. This is because that the index information increase as the number of total OAM-modes increases, which can be verified by the second term on the right-hand of Eq.~\eqref{eq:c_up_re}. Moreover, as $N_{t}$ increase, the probability jammed by other interference users decreases. As $N_{t}$ increases, the gaps of SEs among different $N_{t}$ decreases. The reason is the logarithm function of the number of combinations $K$. For example, the increases are 5.56 bps/Hz at 0 dB from $N_{t}=4$ to 8 and 3.22 bps/Hz from $N_{t}=8$ to 12, respectively. In addition, the SEs increases as channel SNR increases. These results prove that for the given hopped OAM-modes, the increase of total OAM-modes results in the increase of SE and reliability of our proposed scheme in radio vortex wireless communications.

\begin{figure}
\centering
  \includegraphics[width=0.53\textwidth]{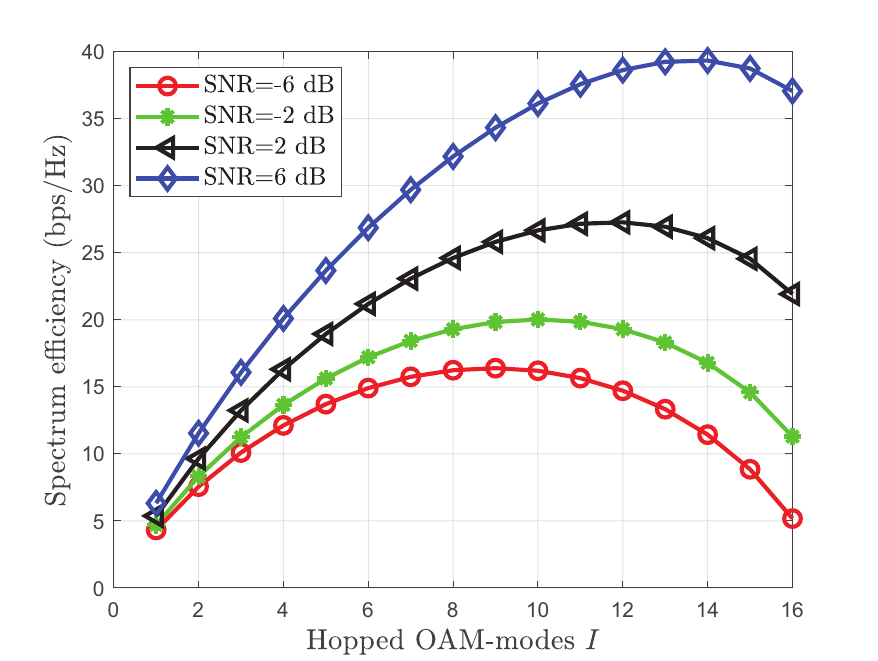}\\
  \caption{The SEs comparisons among different channel SNR versus the hopped OAM-modes.}\label{fig:SE_I_SNR}
\end{figure}

Figure~\ref{fig:SE_I_SNR} compares the SEs of our proposed scheme among different channel SNR versus the number of hopped OAM-modes, where we set $N_{t}=16$, SNR as -6 dB, -2 dB, 2 dB, and 6 dB, respectively. Observing the curve of SE in Fig.~\ref{fig:SE_I_SNR}, the SEs are convex  functions regarding the hopped number $I$ at low channel SNR region. Thus, there exist maximum SEs, which verifies the reasonableness of Section~\ref{sec:How}. The optimal solution of hopped OAM-modes increases as the channel SNR increases. When $I \geq N_{t}/2$, the index information decreases as $I$ increases, while the signal information increases. Thus, there is a trad-off between the signal information and index information at low channel SNR region. As channel SNR increases, index information requires more decrease, thus resulting in more hopped OAM-modes. Fig.~\ref{fig:SE_I_SNR} verifies that our developed scheme can achieve maximum SE at low channel SNR region by making full use of the index information.

\begin{figure}
\centering
  \includegraphics[width=0.53\textwidth]{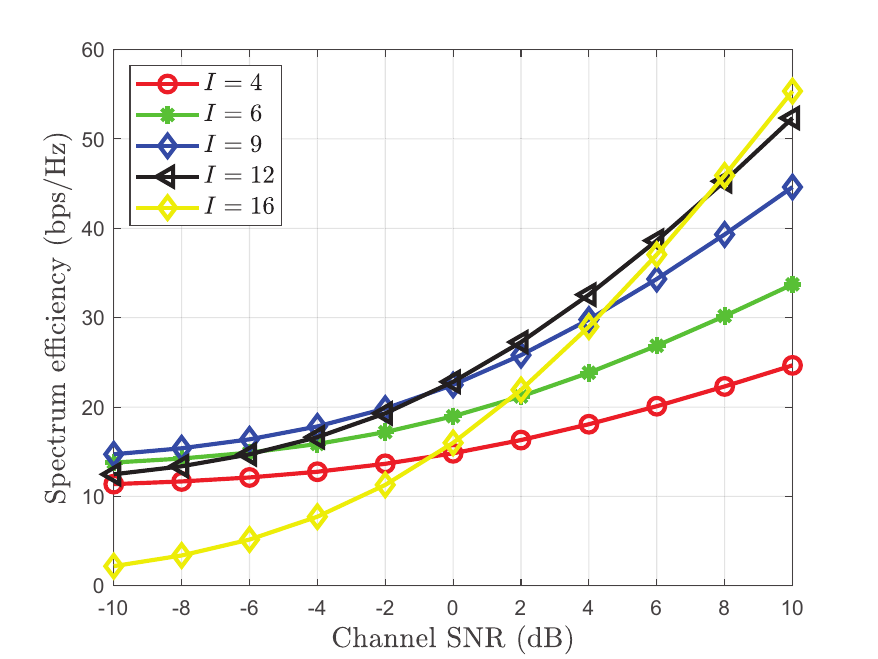}\\
  \caption{The SE comparisons between our proposed scheme and the conventional OAM multiplexing scheme.}\label{fig:SE_snr}
\end{figure}

Figure~\ref{fig:SE_snr} compares the SEs between our proposed scheme and the conventional OAM multiplexing scheme, where we set $N_{t}=16$, $I=4,6,9,12$ and 16, respectively. Clearly, $I=16$ represents the conventional OAM multiplexing scheme. Fig.~\ref{fig:SE_snr} also verifies that there exists maximum SE in low channel SNR region. The SEs increase as the channel SNR increases for a fixed $I$. Our proposed scheme has higher SE than that of the conventional OAM multiplexing scheme in low SNR region. Only in high SNR region, the conventional OAM multiplexing scheme can achieve higher SE. This is because that the signal information plays a major role in SE.

\section{Conclusions} \label{sec:conc}

In this paper, we proposed MH with OAM-based index modulation scheme to achieve high SE while satisfying a certain reliability for wireless communications. Based on the proposed scheme, we derived the upper and lower bounds of SEs and how many hopped OAM-modes should be selected to achieve higher SE in radio vortex wireless communications. Numerical results show that our proposed scheme can achieve the higher SE while satisfying low bit error rate by using the optimal number of hopped OAM-modes in comparison with that using total OAM-modes. Also, the SE increases as the number of total OAM-modes increases for a given number of hopped OAM-modes.

\bibliographystyle{IEEEbib}
\bibliography{References}

\end{document}